\documentclass[journal,comsoc]{IEEEtran}
\usepackage[T1]{fontenc}% optional T1 font encoding

\ifCLASSINFOpdf
\else
\fi
\usepackage{amsmath}
\usepackage{amsthm}
\usepackage{xcolor}
\usepackage{graphicx}
\usepackage{comment}
\usepackage[cmintegrals]{newtxmath}
\newtheorem{lemma}{Lemma}

\begin{document}
\title{Is Dual Polarization Always Optimal?  Unlocking New Spatial Degrees of Freedom }

\author{Adrian~Agustin,~\IEEEmembership{Senior Member,~IEEE,}
        Xavier~Mestre,~\IEEEmembership{Senior Member,~IEEE,}        Carlos~Suárez~García,
        and~Alberto~Pellon~Sancho% <-this % stops a space
%\thanks{M. Shell was with the Department
%of Electrical and Computer Engineering, Georgia Institute of Technology, Atlanta,
\thanks{This work was supported by the WAVE-XL project (PID2024-
160457OB-I00) funded by MCIN/AEI/10.13039/501100011033 and by
FEDER, UE. }}% <-this % stops a space


\markboth{Submitted to IEEE Communications Magazine}{Submitted to IEEE Communications Magazine}
%{Shell \MakeLowercase{\textit{et al.}}: Bare Demo of IEEEtran.cls for IEEE Communications Society Journals}



\maketitle

\begin{abstract}
Current wireless communication systems are designed based on channel models obtained under the assumption of far-field operation, where electromagnetic waves can be approximated by a planar wave. In those circumstances, dual-polarized antennas seem to be the optimal solution for maximizing the spectral efficiency. However, the recent research on XL-MIMO systems represents a paradigm shift, since communication may take place in the near-field and an additional degree of freedom can be achieved. This work presents a modular, low-cost tripolarized antenna at 2.4 GHz band, that is used to experimentally validate theoretical findings about tripolarized antennas, which are commonly based on orthogonal infinitesimal dipoles. The evaluation is carried out in two indoor scenarios and one outdoor scenario, with transmitter comprising multiple tripolarized antennas and a receiver equipped with a single tripolarized antenna. We show that this type of antenna can increase the number of spatial streams up to three or extend the region where terminals are able to achieve two spatial streams.

\end{abstract}


% \fi
\IEEEpeerreviewmaketitle

\section{Introduction}
\IEEEPARstart{M}{ulti}-antenna systems have been the cornerstone of the evolution on the wireless communications over the last 30 years, thanks to the increase in available spatial degrees of freedom and/or diversity for communication. The benefits of multi-antenna systems can improve the system throughput by enabling additional streams or increasing the received signal-to-noise ratio  (SNR). The latter benefit becomes relevant when communications are carried out at higher frequency bands, \cite{Shafi18} or when the channel can be assumed to operate under line-of-sight (LoS) conditions.

Although, even in LoS conditions, additional spatial streams can be transmitted if the transmit and receive antennas satisfy the criterion defined in \cite{Torkildson2011}, wireless communication systems have been evolved toward increasing the number of antennas, i.e. \textit{massive MIMO} systems or \textit{holographic MIMO}, \cite{Sanguinetti23}, \cite{Dardari21}. By increasing the number of antennas or the \textit{antenna aperture length}, it is expected that spatial resolution will improve, yielding higher beamforming gain and spatial multiplexing capability. Nevertheless, as the \textit{aperture length} increases, the conventional far-field assumption used to model the wireless channel and design current wireless systems may no longer hold.

The far-field assumption considers that an electromagnetic (EM) wave can be approximated by a planar wave. However, the new scenario requires working with the actual spherical wave, \cite{Sanguinetti23}, and the channel models for designing wireless communications should properly account for it. In this regard, an analytical model such as the one presented in \cite{Poon05} is physically consistent, as it follows the principles of \textit{antenna theory}, derived from Maxwell's equations, and uses infinitesimal dipoles. A consequence of using a physically consistent channel model that accounts for near-field propagation is that spatial filtering enables not only conventional \textit{beamforming}, which steers the beam towards a given direction, but also \textit{beamfocusing}, which concentrates the energy at a specific location \cite{bjornson2024towards}. The circumstances under which \textit{beamfocusing} is feasible, and how the corresponding region can be characterized, have been investigated in \cite{Mestre25}.

On the other hand, the spectral efficiency of wireless communications can be enhanced by employing antennas with multiple orthogonal polarizations. This is consistent with the dyadic Green's function channel model, which accounts for polarization through the use of orthogonally oriented dipoles:  $\mathrm{pol1}, \mathrm{pol2}, \mathrm{pol3}$, respectively. According to this channel model, the radiative electromagnetic field does not propagate when the polarization is aligned with the propagation direction, which explains why two polarizations ($\mathrm{pol1}, \mathrm{pol2}$) can be effectively exploited in the far-field. In the \textit{near}- or \textit{intermediate}-field regions, however, each antenna in an array is effectively observed from a different propagation direction, potentially enabling the exploitation of this third polarization. %The work in \cite{Poon05} concludes that two polarizations ($\mathrm{pol1}, \mathrm{pol2}$) can be effectively exploited in the far-field, even in LoS conditions, while the third polarization ($\mathrm{pol3}$) may have significant influence in the \textit{near}- or \textit{intermediate}-field regions. 

Tripolarized antennas have been investigated in the literature, see for example \cite{yang25,Wei23}, but experimental validation remains limited. To the best of the authors' knowledge, the first experimental validation of these additional degrees of freedom was presented in \cite{Andrews01}, where two tripolarized antennas were considered in a scattering medium, i.e., a single-tripolarized antenna at transmitter and another at the receiver side, respectively. Similarly, indoor measurements at 2.45 GHz were carried out in \cite{Gutierrez12} using a design based on three orthogonally aligned, center-fed dipole radiators, each 5.4 cm long. With antenna separations of 1 m and 2 m under LoS and non-LOS conditions, i.e. with scattering, the authors evaluate the classic parameters related to polarization, correlation and time dispersion.

Unlike prior work on tripolarized XL-MIMO, this work presents a practical modular antenna and experimentally validates the main results obtained in \cite{Agustin26}, \cite{mestre25ojsp} showing that the number of available spatial degrees of freedom depends on the \textit{aperture length} and the relative position of transmitter and receiver. The key contributions presented in this paper are as follows:
\begin{itemize}
    \item We present a modular and low-cost design for tripolarized antennas based on a combination of patch antenna ($\mathrm{pol1}, \mathrm{pol2}$) and a monopole ($\mathrm{pol3}$), in contrast to the designs adopted in \cite{Andrews01}, \cite{Gutierrez12}.
    \item We perform indoor and outdoor measurements at 2.45 GHz to evaluate the number of available spatial degrees of freedom and to asses how well they match the analytical characterization in \cite{mestre25ojsp}. 
    \item We provide access to estimated wireless channel matrices from the collected measurements at \cite{agustin26dataset}.
\end{itemize}

\section{Analytical Characterization}
\label{sec:theoretical_characterization}

\begin{figure}[t]
\centerline{\includegraphics[width=3 in, clip=true, trim={0 0 0 0}]{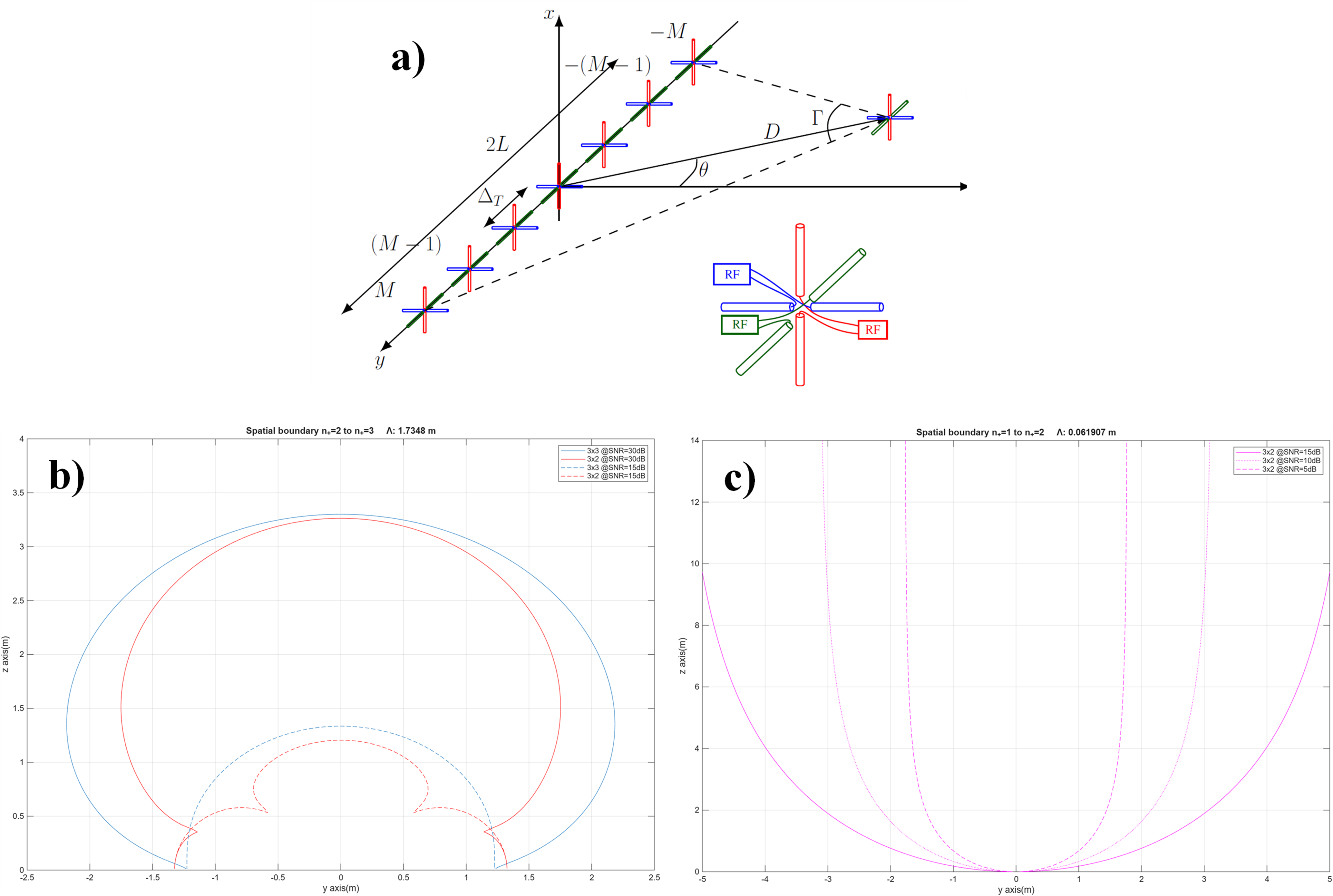}}
\caption{a) Uniform Linear Array with $2M\!\!+\!\!1$ elements and \textit{aperture length} $\Lambda$, where each element has 3 orthogonal infinitesimal dipoles. Receiver at distance $D$ on the \textit{yz}-plane. b) Boundary region for $n_{+}\!\!=\!\!1\!\rightarrow\!2$ under $t_{\textrm{pol}}\!\times\! r_{\textrm{pol}}\!=\!\!2\!\times\!3$. c)  b) Boundary region for $n_{+}\!\!=\!\!2\!\rightarrow\! 3$.}
\label{fig:analytical_scenario}
\end{figure}

The scenario considered for analytical characterization developed in \cite{mestre25ojsp} is presented in Fig. \ref{fig:analytical_scenario}-a). It is assumed to be a transmitting uniform linear array (ULA) with $2M+1$ elements separated by a distance $\Delta_T$ meters (m), deployed along the \textit{y}-axis with an \textit{aperture length} of $\Lambda$ (m). Each element of the array consists of three orthogonal dipoles, each fed by different radio-frequency chain and oriented as follows: $\mathrm{pol1}$ along the \textit{x}-axis, $\mathrm{pol2}$ along the \textit{y}-axis, and $\mathrm{pol3}$ along the \textit{z}-axis. Similarly, the receiver is equipped with a single tripolarized antenna and is assumed to be at distance $D>0$ and \textit{elevation} $\theta\in(-\pi/2,\pi/2)$, under LoS conditions. The number of employed polarizations at the transmitter and receiver sides will be denoted by $t_{\textrm{pol}}$ and $r_{\textrm{pol}}$, respectively. %Taking into account that the SNR is denoted by $\rho_0$ and the number of spatial streams are defined by $n_{+}$, we present the following lemma as summary of the results derived in \cite{mestre25ojsp}.

Since the receiver consists of three dipoles, the communication channel can support up to three spatial degrees of freedom, denoted by $n_+ \in {1,2,3}$. For a fixed transmit power, three spatial streams ($n_+=3$) are typically available in the vicinity of the ULA, whereas only one ($n_+=1$) is available sufficiently far from the array. Figs. \ref{fig:analytical_scenario}-b)-c) illustrate the extent of these regions for different ULA sizes and SNR values, considering $t_\mathrm{pol}=2$ and $t_\mathrm{pol}=3$. More specifically, Fig. \ref{fig:analytical_scenario}-b) shows the boundaries separating the regions where $n_+=2$ and $n_+=3$ spatial DoF are available for a ULA with $M=8$ (17 tripolarized antennas), over an area of $8 \textrm{m}\times5\textrm{m}$. The boundaries are shown for the configurations $t_{\textrm{pol}}\times r_{\textrm{pol}}=2\times3$ (red lines) and $t_{\textrm{pol}}\times r_{\textrm{pol}}=3\times3$ (blue lines), at different transmit powers (normalized SNR values). The results show that the region in which $n_+=3$ spatial DoF are available expands when all antennas operate with three polarizations.

On the other hand, Fig. \ref{fig:analytical_scenario}-c) considers a small ULA with $M=1$ (three tripolarized antennas) and shows the boundaries separating the regions where $n_{+}=1$ and $n_{+}=2$ spatial DoF are available over an area of $5 \textrm{m}\times14\textrm{m}$, for different transmit powers (normalized SNR values). In this case, two spatial DoF ($n_+=2$) are available throughout the considered area when $t_\mathrm{pol}=3$, and hence, no boundary is observed. Such a boundary exists only for the configuration $t_{\textrm{pol}}\times r_{\textrm{pol}}=2\times3$. Note that the availability of two spatial DoF ($n_+=2$) is confined to a region around the broadside direction of the ULA.
%, in this situation $n_{+}=3$ cannot be achieved because the small ULA operates in far-field regime, whereas 
%$n_{+}=2$ are only available in a region around the broadside direction of the ULA. %In contrast, when $t_{\textrm{pol}}\times r_{\textrm{pol}}=3\times3$, $n_{+}=2$ always can be attained according to Lemma \ref{lem:lemma_DoF}.

The boundaries in Fig. \ref{fig:analytical_scenario}-b) and \ref{fig:analytical_scenario}-c) can be analytically described as it is shown in \cite{mestre25ojsp}. We summarize the main results in the following lemma. 
%Taking into account that the SNR is denoted by $\rho_0$ and the number of spatial streams are defined by $n_{+}$, we present the following lemma as summary of the results derived in \cite{mestre25ojsp}.

\begin{lemma}
\label{lem:lemma_DoF}
Let $\rho_0$ denote a $D^2$-normalized version of the SNR, essentially proportional to the transmit power (see \cite{mestre25ojsp} for the exact definition). The number of available spatial DoFs $n_{+}$ and the corresponding availability regions depend on the transmit polarization configuration as follows. 

When $t_\mathrm{pol}=3,r_\mathrm{pol}=3$:
\begin{itemize}
    \item {$n_+>1$ is guaranteed everywhere} if $\rho_{0}>\Lambda^2/12$
    \item The distance at which the transition between $n_+=2$ and $n_+=3$ occurs can be asymptotically approximated by
\begin{align}
D_{\mathrm{th}}^{(2)}\!\!=\frac{\Lambda}{2}\sqrt{A_{1}\left(  \theta\right)
\!+\!\sqrt{A_{1}^{2}\left(  \theta\right) \! - \! A_{2}\left(  \theta\right)
\!+\!\frac{2\cos^{2}\theta}{3\Lambda^2}\rho_{0}}} 
\end{align}
\end{itemize}
with $A_1\left(\theta \right)=3/2 - 29/15\cos^2\theta$, $A_2\left(\theta \right)=2 - 8/3\cos^2\theta$. 

When $t_\mathrm{pol}=2,r_\mathrm{pol}=3$:
\begin{itemize}
    \item    If $\theta=0$, 
    %$\overline{\mathsf{SNR}}_{0}^{(1)}$ converges to a constant $\pi/6$ as $D/L\rightarrow \infty$. This means that at least 
    $n_+ = 2$ modes are always available. % along the broadside of the ULA.
\item For $\theta\neq0$, we can asymptotically describe the distances at which the two transitions occur as 
\begin{align}
D_{\mathrm{th}}^{(1)} &  =\frac{\Lambda}{2}\sqrt{\frac{4\cos^{4}\theta
-16\cos^{2}\theta+10}{3\sin^{2}\theta}+\frac{4\rho_{0}}{\Lambda^2\tan^{2}\theta}} \\
D _{\mathrm{th}}^{(2)} &  = \frac{\Lambda}{2}\sqrt{ B_{1}(\theta)+\sqrt{B_{1}^{2}\left(  \theta\right)  -B_{2}\left(  \theta\right)+\frac{2\cos^{4}\theta}{3\Lambda^2}\rho_{0}}} 
\end{align}
with $B_1\left(\theta \right)=1/12\cos^4\theta - 59/20\cos^2\theta + 64/30$, 
%$C_{1}\left(\theta \right)=\cos^8\theta/144 - 97/360\cos^6\theta +  64391/25200\cos^4\theta - 57/175\cos^2\theta - 422/225$
$B_2\left(\theta \right)=482/75-2/9\cos^6\theta+1138/175\cos^4\theta-6437/525\cos^2\theta $.
\end{itemize}
\end{lemma}
Having introduced the main analytical results, we now turn to their experimental validation.

\begin{figure*}[t]
\centering
 \includegraphics[width=0.65\textwidth]{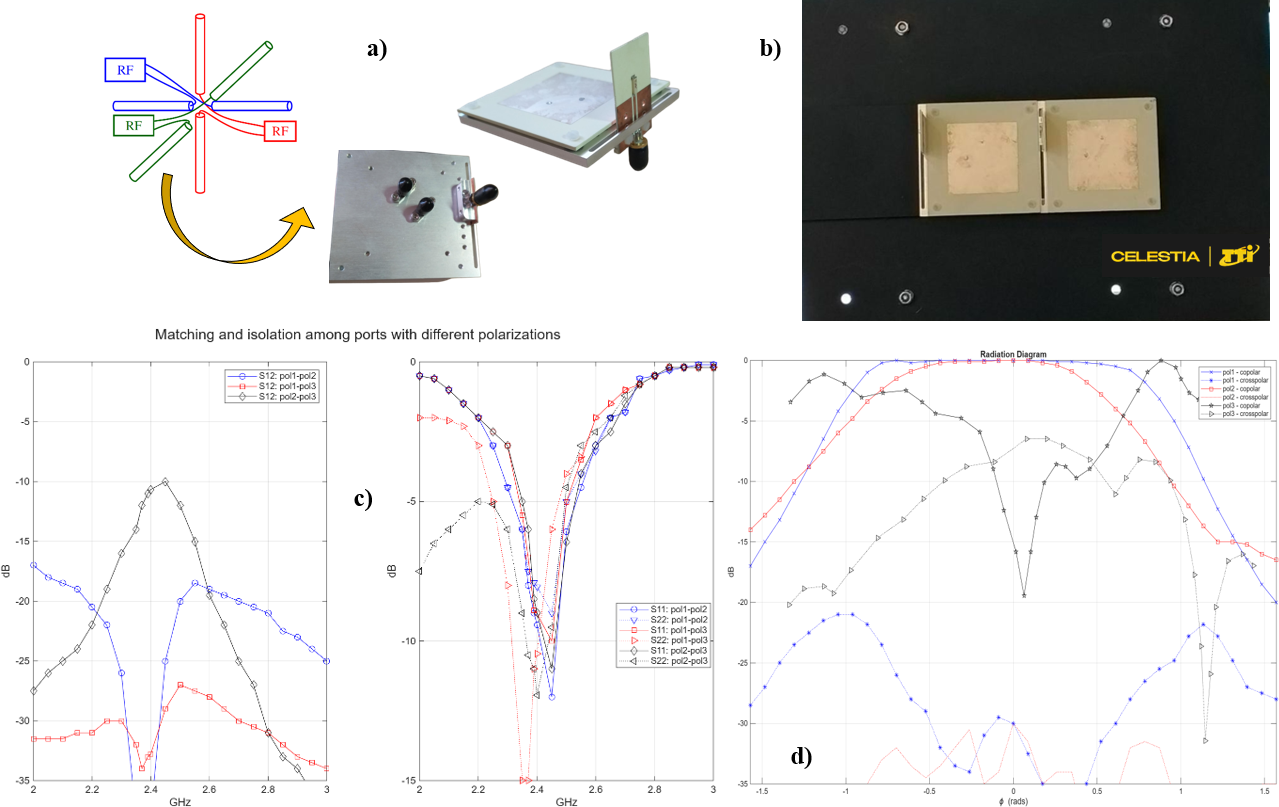}
\caption{a) Patch and monopole antennas. b) Structure to support both type of antennas. c) $2\times1$ Tripolarized antenna. d) \textit{Matching} and {isolation} performance indicators for tripolarized antenna. e) Radiation diagram for $2\times1$ ULA.}
\label{fig:Antenna_Implementation}
\end{figure*}

\section{Tripolarized Antenna Implementation}

The proposed cost-effective solution is based on a modular design in which each tripolarized antenna element consists of two components in the same aperture: (i) dual-polarized antenna or patch antenna, and (ii) a perpendicular monopole providing the \textit{third polarization}. The fabricated Printed Circuit Board (PCB) for the dual polarized antennas ($\mathrm{pol1}$ and $\mathrm{pol2}$) has dimensions $68.75\times68.75$ mm, with a metallic patch of $43.1\times43.1$ mm. The bottom layer is free of metal, and the PCB includes two non-plated holes to allow the connector to pass through and facilitate soldering from the top, as shown in Fig. \ref{fig:Antenna_Implementation}-a). Regarding the PCB for the monopole ($\mathrm{pol3}$), multiple holes are provided to allow the antenna to be moved along \textit{z}-axis, either closer or farther from the ground plane. Both types of antennas are supported by a single structure, in which  the monopole can be adjusted at different positions along the \textit{x}-axis or, alternatively, multiple monopoles can be used per patch antenna, as shown in Fig. \ref{fig:Antenna_Implementation}-b).

The performance of the tripolarized antenna is evaluated in terms of \textit{matching} and \textit{isolation}, as shown in Fig. \ref{fig:Antenna_Implementation}-c) and measured with a vector network analyzer. The dual polarized antenna achieves a \textit{matching} better than 10 dB at 2.45 GHz and 8 dB across the entire operating band (2.4-2.5 GHz). Likewise, the \textit{isolation} between the two polarizations is at least of 20 dB. When comparing the dual-polarized and monopole ports, the \textit{isolation} between $\mathrm{pol1}$ and $\mathrm{pol3}$ exceeds 20 dB, whereas $\mathrm{pol2}$ and $\mathrm{pol3}$ is around 10.5 dB. Regarding the \textit{matching}, it is approximately 10 dB, similar to the dual-polarized case. In total, 20 tripolarized antennas have been manufactured and distributed across two $2\times1$ arrays, as shown at  Fig. \ref{fig:Antenna_Implementation}-b), and two $8\times1$ ULAs.

Finally, the radiation pattern has been characterized using a \textit{standard-gain horn} (SGH) with single polarization as the \textit{probe} in an anechoic chamber. The antenna under test was rotated in azimuth and elevation to extract the 3D radiation pattern. In a $2\times1$ configuration, such as the one depicted in Fig. \ref{fig:Antenna_Implementation}-c), the radiation pattern obtained with only the dual-polarized antennas is shown at Fig. \ref{fig:Antenna_Implementation}-d), attaining the maximum at $\theta=0^\circ$ with an \textit{isolation} of 20 dB between polarizations. When considering the monopole antennas, the radiation pattern exhibits a null at $\theta=0$, while the maximums are obtained at $\theta=\pm \pi/2$. Notice that the monopole complements the dual-polarized antennas providing additional coverage.

\begin{comment}
\begin{figure}[t]
\centerline{\includegraphics[width=2.9 in, clip=true, trim={0 3 0 3}]{figures/Tripol_antenna.png}}
\caption{To be redrawn or mixed with next}
\label{fig:XX}
\end{figure}

\begin{figure}[t]
\centerline{\includegraphics[width=2.9 in, clip=true, trim={0 3 0 3}]{figures/S_parameter.png}}
\caption{To be redrawn or mixed with previous}
\label{fig:XX}
\end{figure}

\begin{figure}[t]
\centerline{\includegraphics[width=2.9 in, clip=true, trim={0 3 0 3}]{figures/Radiation_pattern.png}}
\caption{To be redrawn or mixed with previous}
\label{fig:XX}
\end{figure}
\end{comment}

\begin{figure*}[t]
\centering
\includegraphics[width=0.62\textwidth]{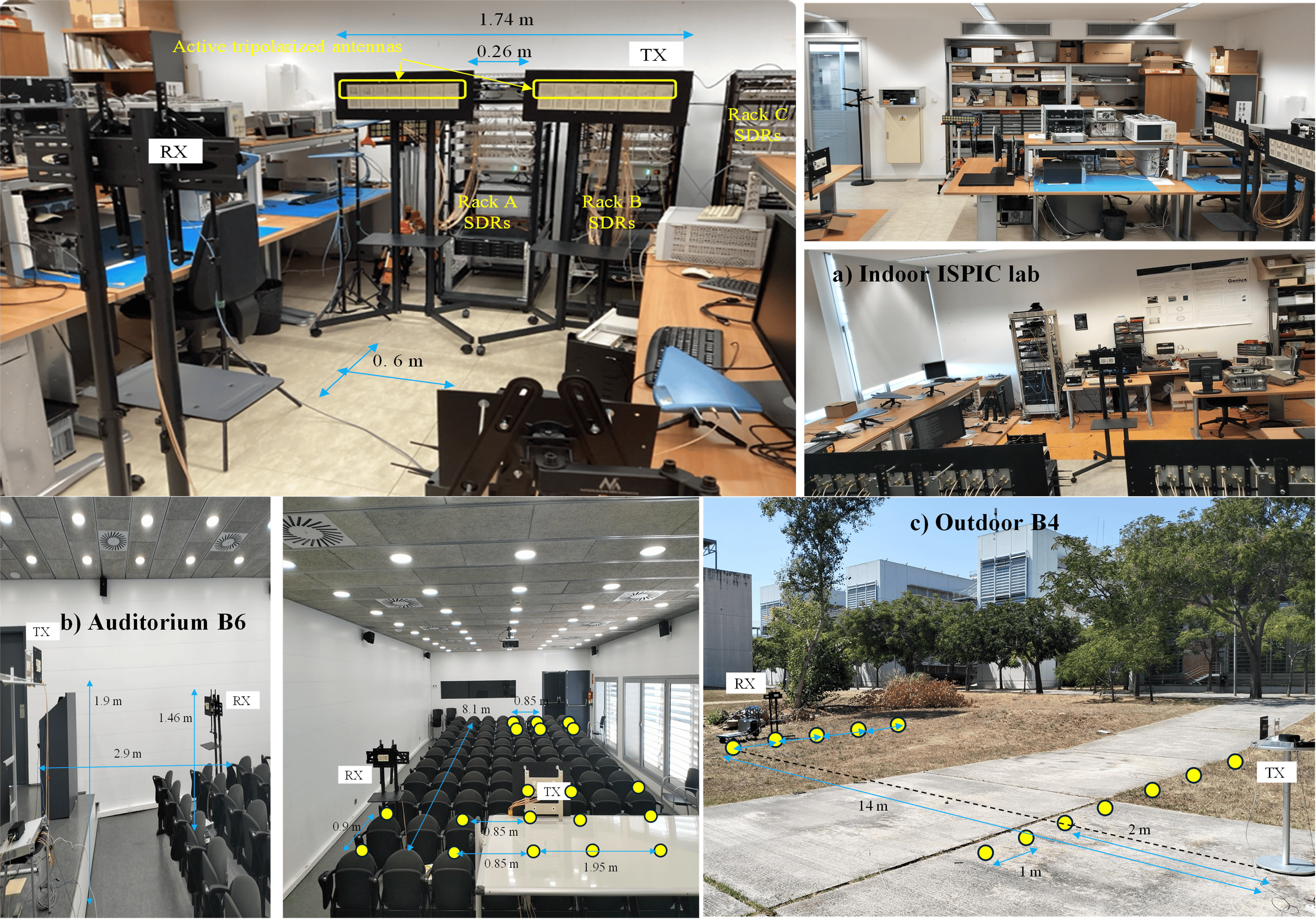}
\caption{ a) Indoor scenario with a transmitter of $\Lambda=1.74$ m (48 RF chains) to show $n_{+}=\{2,3\}$ spatial modes. b) Indoor scenario with a transmitter of $\Lambda=0.06$ m (6 RF chains) to show $n_{+}=\{1,2\}$ spatial modes. c) Outdoor scenario with a transmitter of $\Lambda=0.06$ m  to show $n_{+}=\{1,2\}$ spatial modes.}
\label{fig:experimental_scenario}
\end{figure*}

\section{Experimental Results}
\subsection{Scenario}
The experimental validation is carried out through an \textit{over-the-air} transmission of narrowband signal with 1 MHz bandwidth, where the transmitter assigns a different 2047-length Gold sequence as a pilot to each employed polarization along the array. The receiver estimates the equivalent channel using known pilot sequences, which provides information about the number spatial modes. This is made by the modular laboratory infrastructure deployed at CTTC within the framework of the national project UNICO22 EMERGE6G, that enables the evaluation of new waveforms, integrated communication and sensing techniques and prototype antenna testing. The infrastructure has been specifically designed to validate emerging wireless concepts involving large-scale antenna arrays. The system provides up to 96 radio frequency (RF) chains distributed across 12 Software Defined Radios (SDRs) operating up to 6 GHz and arranged over three racks. All  transceivers, at both the transmitter and receiver side, are synchronized using a 1 Pulse per Second (PPS) signal and 10 MHz reference clock, enabled by White-Rabbit-based devices. More details on the architecture and the transmission set up can be found at \cite{agustin26dataset}.%\cite{agustin26b}.

The validation was carried out in three different scenarios, as shown in Fig. \ref{fig:experimental_scenario}. The evaluation of the regions where up to $n_{+}=3$ spatial modes can be found, as shown in Fig. \ref{fig:analytical_scenario}-c), was conducted in a laboratory with approximate dimensions $4 \textrm{m}\times5\textrm{m}$, as depicted in Fig. \ref{fig:experimental_scenario}-a). The transmitter consists of two $8\times1$ ULAs separated by 0.26 m, yielding an equivalent \textit{aperture length} of $\Lambda=1.74$ m. In total 48 RF chains are used to transmit the signal \textit{over-the-air} when $t_{\textrm{pol}}=3$ or 32 RF chains when $t_{\textrm{pol}}=2$. At the receiver side, 3 RF chains are required when $r_{\textrm{pol}}=3$ or 2 RF chains if $r_{\textrm{pol}}=2$.

In contrast, for the validation of the boundary for $n_{+}\!\!=\!\!1\!\!\rightarrow\!\!2$ we considered a much larger space and reduced \textit{aperture length} to use a total of two active antennas. This selection allows us to compare the results with those presented in \ref{fig:analytical_scenario}-b); otherwise the boundary widens. To this end, we  considered two cases: (i) an indoor scenario, an auditorium with  approximate dimensions $6 \textrm{m}\times15\textrm{m}$ as depicted in Fig. \ref{fig:experimental_scenario}-b), and (ii) an outdoor scenario,  as presented in Fig. \ref{fig:experimental_scenario}-c). Notice that, for this set up, the transmitter  requires only 6 RF chains when $t_{\textrm{pol}}=3$ or 2 RF chains when $t_{\textrm{pol}}=2$  that is, a single SDR, which facilitates the evaluation across multiple scenarios since it not required to move the three racks with the SDRs.
Measurements, which are available at \cite{agustin26dataset}, were obtained with the receiver placed at different positions. At each position, the following combinations were evaluated: $t_{\textrm{pol}}\times r_{\textrm{pol}}=(3\times3), (2\times3), (2\times2)$. In all cases, the patch antenna  polarizations, $\textrm{pol}1$,$\textrm{pol}2$, remained active. The  $t_{\textrm{pol}}\times r_{\textrm{pol}}=(2\times2)$ configuration represents a  conventional dual-polarized wireless system.
\begin{figure*}[t]
\centering
\includegraphics[width=0.85\textwidth]{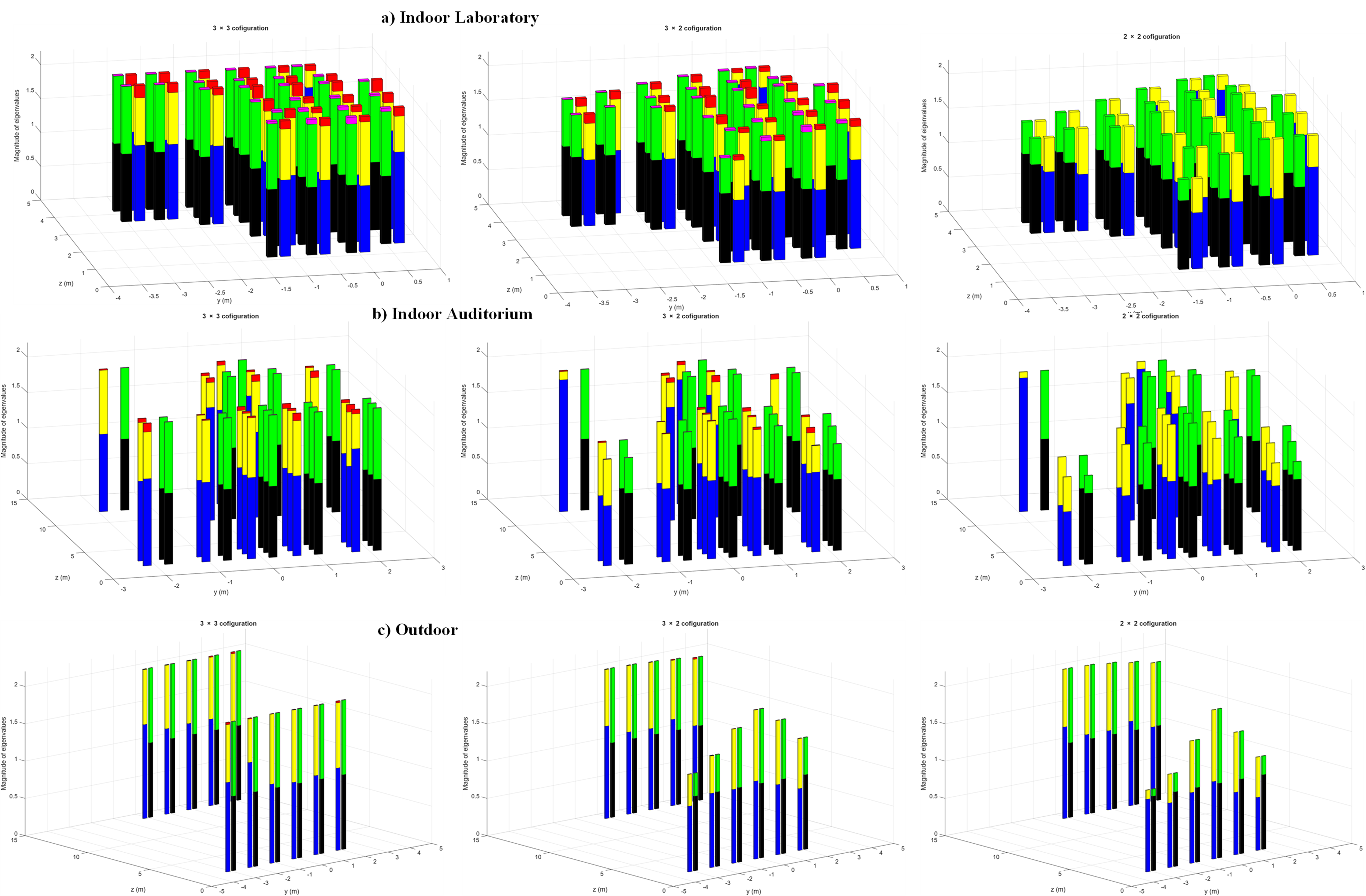}
\caption{Distribution of the eigenvalues of the normalized Gramian of the measured channel  (blue, yellow and red) and LoS channel (black, green, magenta). a) Indoor scenario with $\Lambda=1.74$ m (see Fig. \ref{fig:experimental_scenario}.a). b) Indoor scenario with $\Lambda=0.06$ m (see Fig. \ref{fig:experimental_scenario}.b). c) Outdoor scenario with $\Lambda=0.06$ m (see Fig. \ref{fig:experimental_scenario}.c).}
\label{fig:experimental_eigenvals}
\end{figure*}

\subsection{Performance Indicators}
\begin{figure*}[t]
\centering
\includegraphics[width=0.85\textwidth]{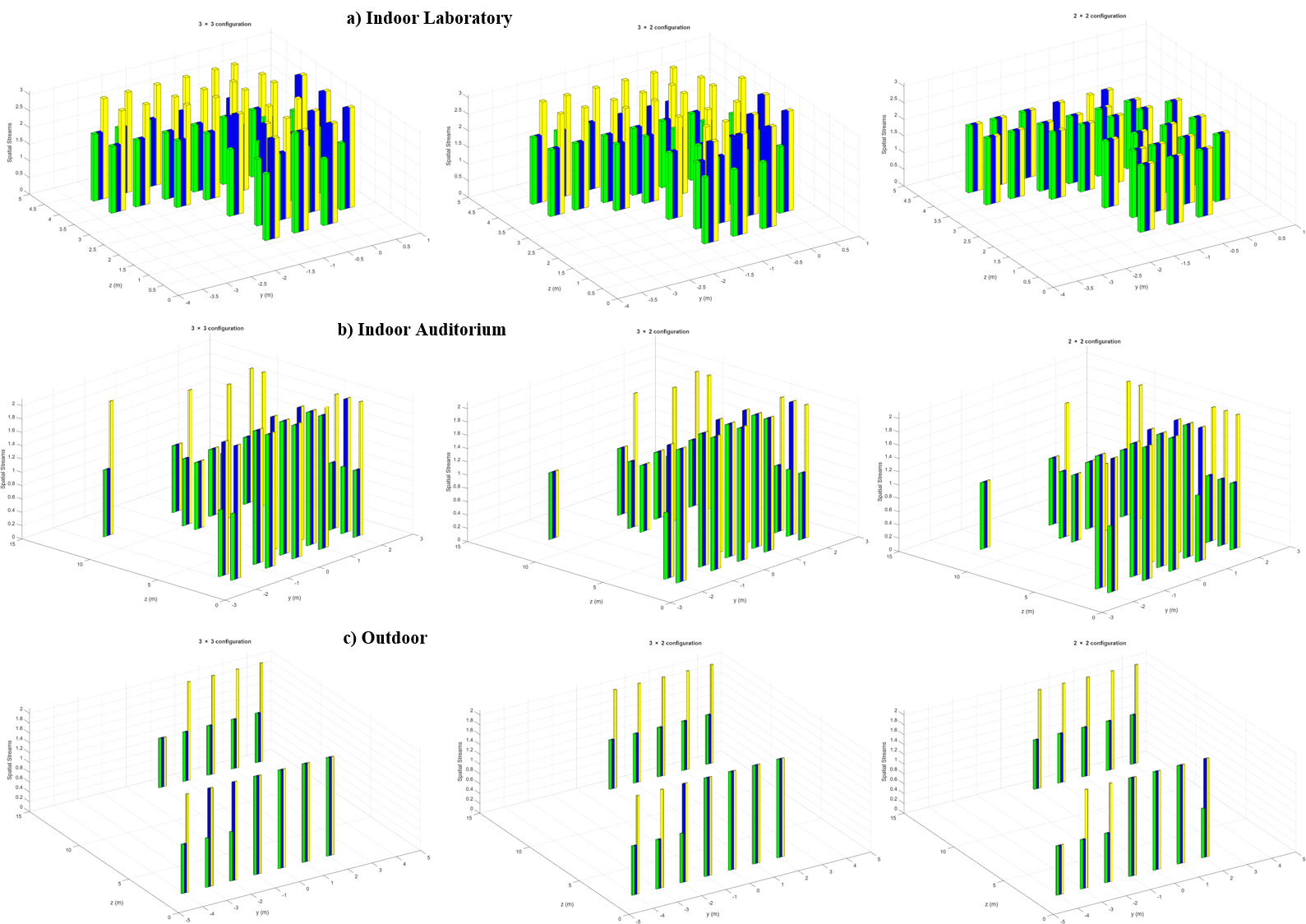}
\caption{Measured spatial streams. a) Indoor scenario with $\Lambda=1.74$ m (see Fig. \ref{fig:experimental_scenario}.a) with  $\rho_0=\{15, 20, 30\} $ dB (green, blue, yellow). b)  Indoor scenario (see Fig. \ref{fig:experimental_scenario}.b) and c) Outdoor scenario (see Fig. \ref{fig:experimental_scenario}.c), both with $\Lambda=0.06$ and  $\rho_0=\{5, 10, 20\} $ dB (green, blue, yellow).}
%b) Indoor scenario with $\Lambda=0.06$ m (see Fig. \ref{fig:experimental_scenario}.b) with  $\rho_0=\{5, 10, 20\} $ dB (green, blue, yellow). c) Outdoor scenario with $\Lambda=0.06$ m (see Fig. \ref{fig:experimental_scenario}.c) with  $\rho_0=\{5, 10, 20\} $ dB (green, blue, yellow).}
\label{fig:experimental_DoF}
\end{figure*}

To compare the experimental results with the theoretical ones under pure LoS conditions, we have considered the \textit{eigenvalues of the normalized Gramian of the channel}, as depicted in  Fig. \ref{fig:experimental_eigenvals}, together with the \textit{number of spatial streams} for the configurations $(3 \times 3)$, $(2\times3)$ and $(2\times2)$, as shown in  Fig. \ref{fig:experimental_DoF} and  Fig. \ref{fig:ideal_DoF}. The experimental measurements have been calibrated at each position and for each configuration by matching the trace of the measured Gramian matrix to that of its ideal counterpart, thereby ensuring the same \textit{spectral sum}, i.e., the same sum of eigenvalues.
%The experimental measurements have been calibrated against the ideal case by matching the power levels, that is, by enforcing the same \textit{trace} value of the Gramian matrices. 

\begin{enumerate}
    \item \textit{Eigenvalues}: The Gramian matrices have at most three eigenvalues, depending on the polarization configuration. In Fig. \ref{fig:experimental_eigenvals}, the colors \textit{blue}, \textit{yellow}, and \textit{red} denote the first, second and third eigenvalue for the experimental Gramian matrix, respectively, while  \textit{black}, \textit{green} and \textit{magenta} denote those of the ideal Gramian matrix. At each point, the magnitude of the three eigenvalues are stacked in a colored bar.
    \item \textit{Spatial streams} ($n_{+}$): The number of active spatial streams is related to the eigenvalues and to the SNR ($\rho_0$), \cite{mestre25ojsp}. Results in Fig. \ref{fig:experimental_DoF} and Fig. \ref{fig:ideal_DoF} illustrate the dependence of $n_{+}$ on $\rho_0$ with different colored bars.
    %show the number of spatial streams for different values of $\rho_0$, represented by colored bars, thereby illustrating the dependence of $n_{+}$ on $\rho_0$.
\end{enumerate}

\subsection{Extending the limits of spatial DoF to $n_{+}=3$ }
According to the theoretical results presented in section \ref{sec:theoretical_characterization}, the experimental scenario depicted in Fig. \ref{fig:experimental_scenario}.a) has the appropriate dimensions and aperture length to achieve $n_{+}=3$ spatial streams with  $(3 \times 3)$, $(2\times3)$ polarization configurations.  Fig. \ref{fig:experimental_eigenvals}-a) presents the magnitude of the eigenvalues at the different measured positions, comparing the measured and ideal cases. The results reveal the following key observations:

\begin{itemize}
    \item \textbf{O1}: The experimental eigenvalue distribution differs slightly  from the theoretical one. The first experimental eigenvalue (\textit{blue}) is higher than the first theoretical eigenvalue (\textit{black}), and a similar trend is observed for the third eigenvalue, \textit{red} and \textit{magenta}, respectively.
    \item \textbf{O2}: In the $(2\times3)$ and $(2\times2)$ configurations, the \textit{spectral sum}, i.e. sum of eigenvalues, remains approximately constant at positions along the antenna broadside, located at (0,0,0). Away from broadside, it decreases, while the first eigenvalue remains larger than the second. By contrast, the $(3\times3)$ configuration maintains a constant \textit{spectral sum} across all positions.
\end{itemize}

\begin{figure*}[t]
\centering
\includegraphics[width=0.85\textwidth]{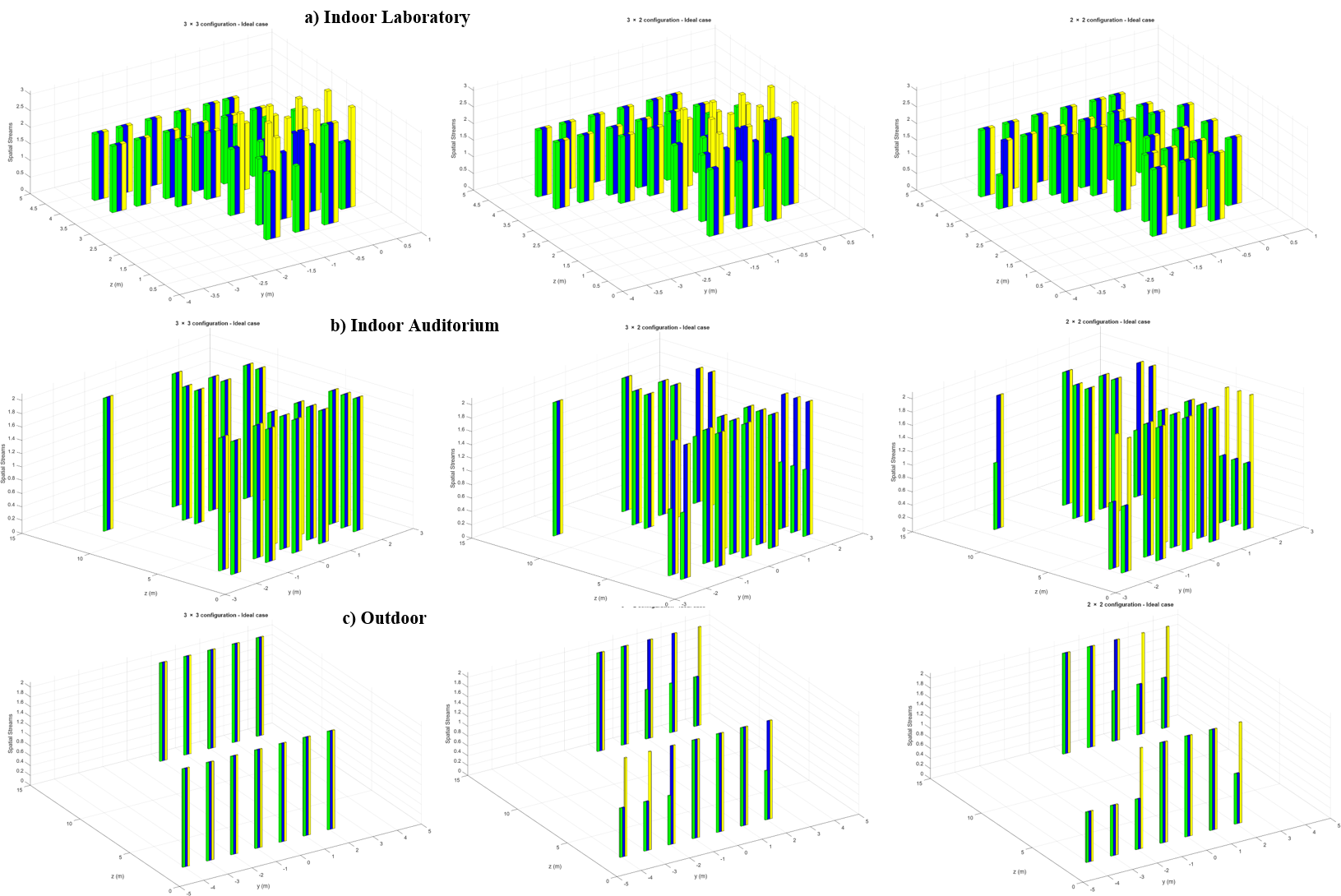}
\caption{Ideal spatial streams. a) Indoor scenario with $\Lambda=1.74$ m (see Fig. \ref{fig:experimental_scenario}.a) with  $\rho_0=\{15, 20, 30\} $ dB (green, blue, yellow). b)  Indoor scenario (see Fig. \ref{fig:experimental_scenario}.b) and c) Outdoor scenario (see Fig. \ref{fig:experimental_scenario}.c), both with $\Lambda=0.06$ and  $\rho_0=\{5, 10, 20\} $ dB (green, blue, yellow).}
%Number of ideal spatial streams in LoS. a) Indoor scenario with $\Lambda=1.74$ m (see Fig. \ref{fig:experimental_scenario}.a) with  $\rho_0=\{15, 20, 30\} $ dB (green, blue, yellow). b) Indoor scenario with $\Lambda=0.06$ m (see Fig. \ref{fig:experimental_scenario}.b) with  $\rho_0=\{5, 10, 20\} $ dB (green, blue, yellow). c) Outdoor scenario with $\Lambda=0.06$ m (see Fig. \ref{fig:experimental_scenario}.c) with  $\rho_0=\{5, 10, 20\} $ dB (green, blue, yellow).}
\label{fig:ideal_DoF}
\end{figure*}

Previous observations are essential for explaining the spatial streams shown in Fig. \ref{fig:experimental_DoF}a). As expected, $(3\times3)$ configuration attains $n_{+}=3$ over a larger region than the $(2\times3)$ configuration at $\rho_0=20$ dB (blue bars). The theoretical characterization identifies closed regions in which the different spatial streams can observed, see Fig. \ref{fig:ideal_DoF}. However, Fig. \ref{fig:experimental_DoF}a) shows that points around (-3,4) and (1,3), both configurations get $n_{+}=3$, whereas nearby points just get $n_{+}=2$. Notice that the magnitude of the ideal third eigenvalue for those positions is zero, while the experimental measurements is larger than zero. These results suggest that the scenario is not purely LoS with some \textit{scattering}, increasing the number of spatial streams. In should be also noted that these positions are close to the large glass window and table with equipment.

\subsection{Expanding the regime for $n_{+}=2$}
Section \ref{sec:theoretical_characterization} has shown that use of tripolarized antennas is beneficial for extending the area where at least $n_{+}=2$ spatial streams are achievable. If tripolarized antennas are not employed simultaneously at transmitter and receiver, the region supporting $n_{+}=2$ is concentrated around the antenna broadside, and its width depends on  \textit{aperture length} and $\rho_0$. Here, we study the case of a small \textit{aperture length} of $\Lambda=0.06$ m in an indoor and outdoor scenarios introduced in Fig. \ref{fig:experimental_scenario}b) and Fig. \ref{fig:experimental_scenario}c), respectively. 

The obtained results for the \textit{spectral sum} are presented in Fig. \ref{fig:experimental_eigenvals}b) and Fig. \ref{fig:experimental_eigenvals}c). Observations \textbf{O1} and \textbf{O2} also apply here. In particular, the experimental measurements in the indoor case exhibit a non-zero third eigenvalue, but this does not translate into an increase in the number of spatial streams for the SNR of interest, see Fig. \ref{fig:experimental_DoF}b). An additional key observation in this case is the following:
\begin{itemize}
    \item \textbf{O3}: For some terminal positions the $(3\times3)$ configuration attains $n_{+}=1$ spatial stream.
\end{itemize}

Although the number of spatial streams obtained with $(3\times3)$ is always greater than or equal those attained with $(2\times3)$ and $(2\times2)$, as expected, in certain cases for indoor and outdoor scenarios only $n_{+}=1$ can be attained, mainly when terminal is not in the broadside or far from the transmitting antenna. This result appears to contradict Lemma \ref{lem:lemma_DoF}, which states that $n_{+}=2$ is achievable everywhere. However, this is due to the fact that even in the considered outdoor scenario in there is some \textit{scattering}. Specifically,   reflections from  the concrete floor in the outdoor scenario, see Fig. \ref{fig:experimental_scenario}c), affect the magnitudes of the first and second eigenvalues. Notice that, in the ideal scenario, the first and second eigenvalues under $(3\times3)$ configuration attain the same value. This is also observed in the experimental eigenvalues when the terminal is placed close to the transmitter, for example at (0,2) in Fig. \ref{fig:experimental_eigenvals}c). However, the values differ when terminal is located at the edge, i.e. at (-4,2), leading to the selection of a single spatial stream.

\section{Open Research lines}
This work presents a preliminary experimental validation of  new spatial degrees of freedom under real-world conditions. However, further research is required along several directions
\begin{itemize}
    \item \textbf{Antenna manufacturing}: The prototype antenna design has demonstrated satisfactory performance for the experimental validations. Further iterations could improve \textit{matching} and \textit{isolation} associated with $\mathrm{pol3}$, while improving the available bandwidth, compactness, and scalability to higher frequency operation.
    \item \textbf{Polarization reconfigurability}: Increasing the number of available orthogonal polarization ports enables active polarization control, adapting the radiating field to propagation environment and variations in antenna orientations. Further research is needed to develop techniques that reduce the complexity of XL-MIMO antenna arrays and potentiality decrease the number of RF chains, \cite{Ding25}. 
    \item \textbf{Impact of scattering}: Although it is known that the scattering can help to increase the number of available degrees of freedom, analytical tractable frameworks for quantifying its impact on the XL-MIMO systems remain limited. Further research is needed to understand how the spatial degrees of freedom evolve in scenarios with a finite number scatterers, including environments with  Reconfigurable Intelligent Surfaces (RIS).
\end{itemize}

\section{Conclusion}
Tripolarized antennas have become an essential tool for improving the performance of wireless communications systems, as they can extend the region where $n_{+}=2$ spatial modes are available and even achieve $n_{+}=3$ spatial modes when the transmitter has an adequate \textit{aperture length}, according the existing results in the literature under LoS conditions. Since the literature often uses orthogonal infinitesimal dipoles to simplify the analytical problem, we present a modular linear array based on a low-cost and simple tripolarized antenna design that combines a patch antenna with two polarizations and a monopole with one polarization, operating in the 2.4 GHz band. This configuration is shown to achieve acceptable \textit{matching} and \textit{isolation} for wireless communications. The new antennas have been evaluated in three different scenarios: two indoor and one outdoor, revealing the superior performance of the $t_{\textrm{pol}} \times r_{\textrm{pol}}=(3\times3)$ over the cases  $(2\times3)$, and $(2\times 2)$, where the later corresponds to the conventional dual-polarized antenna based on patch antennas. The evaluation in the 2.4 GHz band has revealed some discrepancies with the theoretical results due the existence of small \textit{scattering} components, which would disappear when communication is carried out in pure LoS channel, for example by increasing the carrier frequency. The obtained results opens the door to consider more advanced tripolarized antenna design.

\bibliographystyle{IEEEtran}
\bibliography{holographic.bib}


%\end{IEEEbiography}


%\newpage


\begin{IEEEbiographynophoto}{Dr. Adrian Agustin}
%\begin{figure}[htbp]
%  \centering
%  \includegraphics[width=3cm]{photos/Adrian-Augustin-de-Dios.jpg}
%\end{figure}
 (Senior Member, IEEE) (ORCID 0000-0002-5723-9739) is a Senior Researcher at the Centre Tecnològic de Telecomunicacions de Catalunya (CTTC), at ISPIC Research Unit. He received the M.S. and Ph.D in Telecommunication from Universitat Politècnica de Catalunya (UPC), Barcelona, Spain in 2000 and 2008, respectively. From 2008 to 2019 he was a research associate at the SPCOM group at UPC working in the different research areas of wireless communications. In April 2021, he joined the CTTC. He has participated in the EC funded and training networks and several contracts with industry. He has co-advised 3 PhD Thesis and 4 Master Thesis. He collaborated in 4 book chapters, published around 80 papers in conferences and journals, and participated in 3 patents. He is interested in different research areas for developing next generation wireless communications at L1, L2, with Extremely Large Antenna Arrays (ELAA) or XL-MIMO, Deep-Reinforcement Learning (DRL) and Random Access Networks.
\end{IEEEbiographynophoto}

\begin{IEEEbiographynophoto}{Dr. Xavier Mestre}
(Senior Member, IEEE) (ORCID 0000-0002-5723-9739) is Scientific Coordinator at CTTC. He received his MS and PhD in Electrical Engineering in 1997 and 2002 and a licentiate in Mathematics in 2011 from the Technical University of Catalonia (UPC). After working at UPC, he joined CTTC in 2003. He has participated in 12 European projects and coordinated one of then (EMPhAtiC). He led industrial, ESA, and Spanish government projects. He holds four patents and has authored more than 50 journal papers and more than 100 conference articles. He has received two best-paper awards, supervised five PhD theses, and delivered multiple keynotes. His service includes editorial roles with IEEE Transactions on Signal Processing, IEEE and EURASIP technical committee memberships, and conference leadership roles, including chairing or co-chairing IEEE ICASSP in 2020 and 2026.
\end{IEEEbiographynophoto}

\begin{IEEEbiographynophoto}{Mr. Carlos Suárez García }
(ORCID 0000-0001-6346-4687) is a Senior Antenna Engineer at CELESTIA TTI, in Santander, Spain. He received the Bachelor and the M.S in Telecommunication from Universidad de Oviedo (UniOvi), Gijón, Spain in 2018 and 2020, respectively. From 2021 up to now, he has worked as an antenna engineer for different applications, including SatCom, radar, defense and mobile communications.
\end{IEEEbiographynophoto}

\begin{IEEEbiographynophoto}{Mr. Alberto Pellon Sancho} is Head of the antenna department at CELESTIA TTI, in Santander, Spain. His main areas of expertise include SOTM, manpack, and phased-array satellite terminals, as well as broadband and multibeam antennas for spaceborne and ground gateway applications.
\end{IEEEbiographynophoto}


%\vfill


\end{document}